\documentclass[aps,twocolumn,groupedaddress,showpacs,showkeys]{revtex4-1}
\usepackage{newpxtext,newpxmath}

\usepackage{graphics}
\usepackage[unicode=true,pdfusetitle,
 bookmarks=true,bookmarksnumbered=false,bookmarksopen=false,
 breaklinks=false,pdfborder={0 0 0},backref=false,colorlinks=true,citecolor=blue]
 {hyperref}
\usepackage{bbm}
\usepackage{amsmath}
\usepackage[utf8]{inputenc}

\begin{document}

\title{Path Predictability and Quantum Coherence in Multi-Slit Interference}
\thanks{Published: \bf{\large Phys. Scr. (2019).\\ doi:\href{https://doi.org/10.1088/1402-4896/ab1cd4}{10.1088/1402-4896/ab1cd4}}}
\author{Prabuddha Roy}
\affiliation{Department of Physics, National Institute of Technology, Patna, India.}
\email{prabuddharoy.94@gmail.com}
\author{Tabish Qureshi}
\email{tabish@ctp-jamia.res.in}
\affiliation{Centre for Theoretical Physics, Jamia Millia Islamia, New Delhi,
India.}


\begin{abstract}
In an asymmetric multislit interference experiment, a quanton is more
likely to pass through certain slits than some others. In such a situation
one may be able to predict which slit a quanton is more likely to go
through, even without using any path-detecting device. This allows one
to talk of {\em path predictability}. It has been shown earlier that
for a two-slit interference, the predictability and fringe visibility
are constrained by the inequality $\mathcal{P}^2+\mathcal{V}^2\le 1$.
Generalizing this relation to the case of more than two slits is 
still an unsolved problem. A new definition for predictability for
multi-slit interference is introduced. It is shown that this predictability
and {\em quantum coherence} follow a duality relation
$\mathcal{P}^2+\mathcal{C}^2\le 1$, which saturates for all pure states.
For the case of two slits, this relation
reduces to the previously known one.
\end{abstract}

\maketitle

\section{Introduction}

Bohr's principle of complementarity \cite{bohr}, also commonly referred to
as wave-particle duality, has been under research attention for over several
decades. The basic idea is that wave nature and particle nature are two
aspects of a quanton, which cannot be observed at the same time.
If an experiment fully reveals the particle nature of a quanton, it
will completely hide its wave nature, and vice-versa. A two-slit 
interference experiment, carried out with photons or massive particles,
provides an ideal setting to study the principle of complementarity.
In such an experiment, the interference is the obvious signature of the wave
nature of the quanton. If one is able to acquire knowledge about which
of the two slits the quanton went through, in a particular case, one may
say that the quanton behaved like a particle, as it went through only 
one slit and and not through both, as a wave would do.
Starting from Bohr's idea that an experiment can only reveal either wave nature
or particle nature, one can try to go further by asking, can one
{\em partially} observe both wave and particle natures at the same time?
In other words, can an experiment which only partially reveals (say) the
particle nature of a quanton, also partially reveal its wave nature?
In order to address such a question, one has to first introduce quantitative
measures of wave and particle natures. From the preceding discussion,
one may consider the sharpness of interference as a measure of wave nature,
and the amount of knowledge, regarding which slit the quanton went through,
may quantify the particle nature.

The first attempt to state the principle of complementarity in a
quantitative manner was made by
Wootters and Zurek \cite{wootters} where they analyzed the effect of 
introducing a path-detecting device in a two-slit interference experiment.
This work was later extended by Englert who derived a wave-particle duality
relation \cite{englert}
\begin{equation}
{\mathcal D}^2 + {\mathcal V}^2 \le 1,
\label{englert}
\end{equation}
where ${\mathcal D}$ is path distinguishability, a measure of the particle
nature, and ${\mathcal V}$ the visibility of interference, a measure of
wave nature. This inequality is saturated for all pure states.
This kind of wave-particle duality has also been studied using various
other approaches \cite{coles,vaccaro}.

However, there is another way in which the issue of wave-particle duality
has been extensively studied, without introducing a which-way measuring,
or path-detecting device. The idea is that if the two paths, corresponding
to the quanton passing through the two slits, have a priori different 
probabilities, one could predict which of the two paths the quanton might
have taken, with a success which is more than a fifty percent random guess.
This line of thought was pioneered by Greenberger and Yasin \cite{greenberger},
and followed up by Jaeger, Shimoni and Vaidmann \cite{jaeger}, which resulted
in a different kind of duality relation
\begin{equation}
\mathcal{P}^2 + \mathcal{V}^2 \le 1,
\label{GY}
\end{equation}
where $\mathcal{P}$ is a path-predictability, and $\mathcal{V}$ the
visibility of interference. If one can predict the path of the quanton
in a nontrivial manner, one can argue that there is a certain degree of
particle nature associated with the quanton. The fringe visibility
quantifies its wave nature, as in the previous case. The inequality (\ref{GY})
has also been referred to as wave-particle duality relation, but it is
very different from (\ref{englert}), and corresponds to a completely different 
scenario.

The duality relation of the first kind (\ref{englert}) has also been
generalized to the case of interference involving more than two
slits \cite{3slit,cd,nslit,bagan,coles1}.
As regards the duality relation of the second kind (\ref{GY}),
Jaeger, Shimoni and Vaidmann, at the end of their paper \cite{jaeger}, had
proposed that this duality relation, involving path
predictability, should also be extendable to the case of n slits.
There has been substantial effort towards generalizing the duality relation
(\ref{GY}), involving path predictability, to the multi-slit case
\cite{durr,bimonte,englertmb}. Although predictability and visibility may be
defined in various ways, D\"urr \cite{durr} had proposed certain criteria,
on some physical grounds, which any new predictability and visibility
should satisfy. For example, one criteria for any visibility is that it should
be possible to define the visibility based only on the interference pattern,
without referring to the density matrix of the quanton.
A completely satisfactory result for n-slit interference has not been obtained
yet in this regard.
This is the question we have addressed in this work.

\section{Wave-particle duality in n-slit interference}

D\"urr carried out a detailed analysis of predictability and visibility
for multi-path interference, and proposed some general criteria which any
new measure of predictability and visibility should satisfy \cite{durr}.
One notable point of D\"urr's analysis was that he believed that any
new duality relation should become an equality for all pure states.
Englert took a slightly different approach which relaxed the criterion that
the duality relation be saturated for all pure states \cite{englertmb}.

\subsection{Wave nature: generalized visibility}

The first step is to introduced a new generalized visibility which would
be suitable for a n-path interference. It has been argued before that a recently
introduced measure of quantum coherence, in context of quantum information
theory, can be a good measure of wave nature \cite{cd}.
The normalized coherence is defined as \cite{cd}
\begin{equation}
{\mathcal C} \equiv {1\over n-1}\sum_{j\neq k} |\rho_{jk}| ,
\label{C}
\end{equation}
where $\rho_{jk}$ are the matrix elements of the density operator of the
quanton in a particular basis. This coherence is based on the coherence
measure introduced by Baumgratz, Cramer, Plenio in a seminal
paper \cite{coherence}.  In the context of multi-path interference, the
basis states are the states of the quanton corresponding to its passing 
through various slits. It has been further demonstrated that this coherence 
can be experimentally measured in a multi-slit interference experiment
\cite{tania}. Coherence $\mathcal{C}$ also satisfies D\"urr's criteria for
a good visibility \cite{tania}.  This makes coherence $\mathcal{C}$ as a
good candidate for the generalized visibility.
In another development, Bagan et al. \cite{group} connected the wave nature
of a quanton to a property called group asymmetry, which is related to
coherence. Using this they derived a duality relation between path knowledge
of a quanton, and its group asymmetry. For our purpose, we will confine
ourselves to coherence $\mathcal{C}$.

It is interesting to compare $\mathcal{C}$ with the generalized visibility
introduced by Dürr \cite{durr}
\begin{equation}
\mathcal{V} = \left(\frac{n}{n-1}\sum_j\sum_{k\neq j} |\rho_{jk}|^2\right)^{1/2}.
\end{equation}
As one can see, Dürr's visibility is very close to $\mathcal{C}$.
For the case of two slits ($n=2$), both $\mathcal{C}$ and Dürr's $\mathcal{V}$
reduce to the visibility defined by Greenberger and Yasin, i.e., 
$\mathcal{V} = 2|\rho_{12}|$ \cite{greenberger}.


\subsection{Particle nature: path predictability}

Next we move on to defining the path predictability for n-path interference.
We start by looking at Greenberger and Yasin's path predictability for two
slits \cite{jaeger}
\begin{equation}
\mathcal{P} = \sqrt{1 - 4\rho_{11}\rho_{22} }.
\label{PGY}
\end{equation}
It may be noted that if the second term in the square root in the above equation
is written as $(\sum_{j\neq k} \sqrt{\rho_{jj}}\sqrt{\rho_{kk}})^2$, the factor
of 4 emerges naturally.  Taking cue from this, 
we introduce the following path predictability for the case of $n$ paths, 
\begin{equation}
\mathcal{P} \equiv \sqrt{1 - \left({1\over n-1}\sum_{j\neq k} \sqrt{\rho_{jj}}\sqrt{\rho_{kk}} \right)^2},
\label{P}
\end{equation}
where the factor $\frac{1}{n-1}$ has been introduced to normalize the
second term inside the square root. For $n=2$, (\ref{P}) reduces to
(\ref{PGY}).

We list below, D\"urr's criteria for a good predictability \cite{durr}:\\
(1) $\mathcal{P}$ should be a continuous function of the probabilities
$\rho_{jj}$.\\
(2) If we know the quanton’s way for sure (i.e., $\rho_{j j}=1$ for
one beam, implying $\rho_{j j}=0$ for all other beams), $\mathcal{P}$ should
reach its global maximum.\\
(3) If all ways are equally likely (i.e., all $\rho_{jj}=1/n$), $\mathcal{P}$
should reach its global minimum.\\
(4) Any change toward equalization of the probabilities
$\rho_{11}, \rho_{22}, \dots, \rho_{nn}$ should decrease $\mathcal{P}$.
Thus if $\rho_{11} < \rho_{22}$ and
increase $\rho_{11}$, decreasing $\rho_{22}$ an equal amount so that
$\rho_{11}$ and $\rho_{22}$ are more nearly equal, then
$\mathcal{P}$ should decrease.

Now (\ref{P}) represents a continuous function of various probabilities
$\rho_{ii}$. If (say) $\rho_{ii}=1$, and the rest are zero, then (\ref{P})
yields $\mathcal{P}=1$. For $\rho_{ii}=1/n$ for all $i$, (\ref{P})
yields $\mathcal{P}=0$. For the 4th condition, let us assume
$\rho_{11} < \rho_{22}$ and we increase $\sqrt{\rho_{11}}$ by $\epsilon$ and
decrease $\sqrt{\rho_{22}}$ by $\epsilon$.
The changed predictability is
\begin{equation}
\mathcal{P}' = \sqrt{1 - \left({1\over n-1}\sum_{j\neq k} \sqrt{\rho_{jj}}
\sqrt{\rho_{kk}} + 2\epsilon(\sqrt{\rho_{22}}-\sqrt{\rho_{11}}-\epsilon)\right)^2},
\end{equation}
Since $\epsilon$ is assumed to be smaller than $\sqrt{\rho_{22}}-\sqrt{\rho_{11}}$,
$\mathcal{P}' < \mathcal{P}$, and condition 4 is also satisfied by (\ref{P}).
Hence (\ref{P}) satisfies all of D\"urr's criteria for a good predictability.

\subsection{Wave-particle duality}

With our definitions of path predictability and generalized visibility, or
coherence, in place, we move on to finding the bound that the two together
should satisfy. Using (\ref{C}) and (\ref{P}), we can write
\begin{eqnarray}
\mathcal{P}^2 + \mathcal{C}^2 = 1 - \left(\tfrac{1}{ n-1}\sum_{j\neq k}
\sqrt{\rho_{jj}}\sqrt{\rho_{kk}} \right)^2
 + \left(\tfrac{1}{ n-1}\sum_{j\neq k} |\rho_{jk}|\right)^2 .\nonumber\\
\label{pc0}
\end{eqnarray}
Let us concentrate on a principal 2x2 sub-matrix of $\rho$ given by
$\left(\begin{matrix}
\rho_{jj} & \rho_{jk}\\
\rho_{kj} & \rho_{kk}\\
\end{matrix}\right) $.
This matrix is positive semi-definite, and for any such matrix
$|\rho_{jk}| \le \sqrt{\rho_{jj}\rho_{kk}}$ \cite{horn}.
Using this result, we can argue that for the two sums in the rhs of
(\ref{pc0}), the number of terms in each are
the same, and each term in the first sum is greater than or equal
to the corresponding term in the second sum. Consequently, the first sum
is greater than or equal to the second sum. Hence the rhs of
(\ref{pc0}) is less than or equal to 1. Thus we can write the inequality
\begin{eqnarray}
\mathcal{P}^2 + \mathcal{C}^2 \le 1 .
\label{pc}
\end{eqnarray}
This is a new wave-particle duality relation involving path predictability and
quantum coherence.

Let us consider a quanton in a pure state
\begin{equation}
|\psi\rangle = \sum_{j=1}^n c_j |\psi_j\rangle,
\end{equation}
where $|\psi_j\rangle$ are the states corresponding to the particle following
the j'th path. Thus the states $\{|\psi_j\rangle\}$ are mutually orthogonal
and normalized. For this pure states, the density matrix elements are given by
$\rho_{jj} =  |c_j|^2, |\rho_{jk}| =  |c_j||c_k|$. For this case
\begin{equation}
- \left(\tfrac{1}{ n-1}\sum_{j\neq k}
\sqrt{\rho_{jj}}\sqrt{\rho_{kk}} \right)^2
 + \left(\tfrac{1}{ n-1}\sum_{j\neq k} |\rho_{jk}|\right)^2 = 0,
\end{equation}
and the duality relation (\ref{pc}) becomes an equality
\begin{eqnarray}
\mathcal{P}^2 + \mathcal{C}^2 = 1 .
\label{pc1}
\end{eqnarray}
Thus (\ref{pc})  represents a wave-particle duality relation for n-path
interference, which saturates for all pure quanton states.

The expression for predictability (\ref{P}), particularly for a pure
quanton state, looks interestingly similar to the expression for
``path-distinguishability" proposed earlier \cite{nslit}, namely,
$\mathcal{D} = \sqrt{1 - \left({1\over n-1}\sum_{i\neq j} |c_ic_j| |\langle d_i|d_j\rangle|\right)^2}$,
where $c_j$ is the amplitude for the quanton to
pass through the $j$'th slit, and $|d_j\rangle$'s are the normalized states of a
path-detecting device. Although the two are very different experiments,
one invloving an additional path-detecting device, notice that if all
$|d_j\rangle$'s are identical to each other, the expression for 
$\mathcal{D}$ reduces to the expression for $\mathcal{P}$ given by (\ref{P}).
Now, the expression for $\mathcal{D}$ is related to the maximal probability
of {\em unambiguously} discriminating between various $|d_j\rangle$'s.
For all identical $|d_j\rangle$'s such probability becomes meaningless,
and one does not know what meaning can be assigned to $\mathcal{D}$.
Thus, one might be tempted to conclude that the similarity between $\mathcal{D}$
and $\mathcal{P}$ is accidental. However, we feel that there may be 
some deeper physical meaning hidden in this similarity. As of now,
we are unable to draw any more conclusion from this.

\subsection{Specific cases}

Now that we have the general duality relation (\ref{pc}) which holds for
arbitrary number of slits, let us look at some particular cases. As mentioned
earlier too, for $n=2$, $\mathcal{P}=\sqrt{1 - 4\rho_{11}\rho_{22} }$,
and $\mathcal{C}=2|\rho_{12}|$. These expressions agree with the predictability
and visibility of Greenberger and Yasin \cite{greenberger}.
Thus, for two slits, our general duality relation (\ref{pc}) reduces to 
Greenberger and Yasin's duality relation (\ref{GY}).

For three-slit interference ($n=3$), the wave and particle measures take
the following form
\begin{eqnarray}
	{\mathcal C} &=& |\rho_{12}| + |\rho_{23}| + |\rho_{13}| ,\nonumber\\
	{\mathcal P} &=& \sqrt{1-\left(\sqrt{\rho_{11}}\sqrt{\rho_{22}} + \sqrt{\rho_{22}}\sqrt{\rho_{33}}
	+ \sqrt{\rho_{11}}\sqrt{\rho_{33}}\right)^2}.~~~~~~
\label{C}
\end{eqnarray}
The above relations hold true for both pure and mixed density matrices of
the particle, and may be contrasted with similar relations in earlier
works \cite{durr,bimonte,englertmb}.

\section{Conclusion}

We have introduced a new definition of path predictability of a quanton
in a n-path interference. We have shown that this new path predictability
and quantum coherence follow an inequality $\mathcal{P}^2 + \mathcal{C}^2 \le 1$.
This inequality is saturated by all pure quanton states. This should be 
considered a new wave-particle duality relation for n-path interference.
For the special case of two slits, this relation reduces to 
Greenberger and Yasin's well known relation.


\begin{thebibliography}{0}
\bibitem{bohr} N. Bohr, ``The quantum postulate and the recent development of
atomic theory," {\em Nature (London)} {\bf 121}, 580-591 (1928). 

\bibitem{wootters} W. K. Wootters and W. H. Zurek,
``Complementarity in the double-slit experiment: Quantum nonseparability
and a quantitative statement of Bohr's principle",
{\em Phys. Rev. D} {\bf 19}, 473 (1979).

\bibitem{englert} B-G. Englert, ``Fringe visibility and which-way information:
an inequality", {\em Phys. Rev. Lett.} {\bf 77}, 2154 (1996).

\bibitem{coles}  P.J. Coles, J. Kaniewski, S. Wehner, ``Equivalence of wave-particle duality to entropic uncertainty," {\em Nature Communication} {\bf 5}, 5814 (2014).

\bibitem{vaccaro} J.A. Vaccaro, ``Particle-wave duality: a dichotomy between
symmetry and asymmetry,"
{\em Proc. R. Soc. A.} {\bf 468}, 1065-1084 (2012).

\bibitem{greenberger} D.M. Greenberger, A. Yasin,
``Simultaneous wave and particle knowledge in a neutron interferometer",
Phys. Lett. A 128, 391 (1988).

\bibitem{jaeger} G. Jaeger, A. Shimony, L. Vaidman, ``Two interferometric complemenarities," {\em Phys. Rev. A} {\bf 51}, 54 (1995).

\bibitem{3slit}  M.A. Siddiqui, T. Qureshi, ``Three-slit interference:
A duality relation", {\em Prog. Theor. Exp. Phys.} {\bf 2015}, 083A02 (2015).

\bibitem{cd}  M.N. Bera, T. Qureshi, M.A. Siddiqui, A.K. Pati,
``Duality of quantum coherence and path distinguishability", {\em Phys. Rev. A} {\bf 92}, 012118 (2015).

\bibitem{nslit}  T. Qureshi, M.A. Siddiqui, ``Wave-particle duality in N-path interference", {\em Ann. Phys.} {\bf 385}, 598-604 (2017).

\bibitem{bagan} E. Bagan, J.A. Bergou, S.S. Cottrell, M. Hillery, ``Relations between Coherence and Path Information," {\em Phys. Rev. Lett.} {\bf 116}, 160406 
(2016).

\bibitem{coles1}  P.J. Coles, ``Entropic framework for wave-particle duality in multipath interferometers," {\em Phys. Rev. A} {\bf 93}, 062111 (2016).

\bibitem{durr} S. D\"{u}rr, ``Quantitative wave-particle duality in multibeam interferometers," {\em Phys. Rev. A} {\bf 64}, 042113 (2001).

\bibitem{bimonte} G. Bimonte, R. Musto, ``Comment on `Quantitative wave-particle duality in multibeam interferometers'," {\em Phys. Rev. A} {\bf 67}, 066101
 (2003).

\bibitem{englertmb} B-G. Englert, D. Kaszlikowski, L.C. Kwek, W.H. Chee, ``Wave-particle duality in multi-path interferometers: General concepts and three-path interferometers," {\em Int. J. Quantum Inform.} {\bf 6}, 129 (2008).

\bibitem{coherence} T. Baumgratz, M. Cramer, M.B. Plenio,
``Quantifying Coherence",
{\em Phys. Rev. Lett.} {\bf 113}, 140401 (2014).

\bibitem{tania} T. Paul, T. Qureshi, ``Measuring quantum coherence in multi-slit interference,"
 {\em Phys. Rev. A} {\bf 95}, 042110 (2017).

\bibitem{group} E. Bagan, J. Calsamiglia, J.A. Bergou, M. Hillery, ``A generalized wave-particle duality relation for finite groups,"
 {\em J. Phys. A: Math. Theor.} {\bf 51}, 414015 (2018).

\bibitem{horn} R.A. Horn, C.R. Johnson, {\em Matrix Analysis} (Cambridge University Press, Cambridge, 1985), p. 398.


\end{thebibliography}
\end{document}